\documentclass[12pt]{article}
\usepackage{amsmath}
\usepackage{epsfig}
\usepackage{amssymb}
\usepackage{cite}
\input{epsf}
\def\beq{\begin{equation}}
\def\eeq{\end{equation}}
\def\beqa{\begin{eqnarray}}
\def\eeqa{\end{eqnarray}}

\def\beq{\begin{equation}}
\def\eeq{\end{equation}}
\def\beeq{\begin{eqnarray}}
\def\eeeq{\end{eqnarray}}
\def\to{\rightarrow}

\newcommand{\sign}{\text{sign}}

\def\b0{b_0}

\begin{document}

\DeclareGraphicsExtensions{.pdf,.gif,.jpg,.eps,.ps,.epsi}

\begin{titlepage}
\renewcommand{\thefootnote}{\fnsymbol{footnote}}
\begin{flushright}
ICAS 027/17 \\
     \end{flushright}
\par \vspace{10mm}
\begin{center}
{\large \bf Transversely polarized Drell-Yan asymmetry $A_{TT}$  at NLO}
\end{center}
\par \vspace{2mm}
\begin{center}
{\bf Daniel de Florian}

\vspace{5mm}
International Center for Advanced Studies (ICAS), ECyT-UNSAM, \\
Campus Miguelete, 25 de Mayo y Francia, (1650) Buenos Aires, Argentina \\
 \vspace{3mm}
and \\
 \vspace{3mm}
 Institute for Theoretical Physics, T\"ubingen University,
Auf der Morgenstelle 14, 72076 T\"ubingen, Germany

\end{center}


\par \vspace{9mm}
\begin{center} {\large \bf Abstract} \end{center}
\begin{quote}
\pretolerance 10000
We present the first fully differential next-to-leading order QCD calculation for lepton production
  in transversely polarized hadronic collisions, $p\uparrow p\uparrow \to \ell^{\pm} X$, where the lepton arises from the decay of an electroweak gauge boson. The calculation is implemented in the Monte-Carlo like code `CHE' 
 that already includes the unpolarized and longitudinally polarized cross sections
and may be readily used to perform a comparison to experimental data and to extract information on the related parton distributions.
We analyze the perturbative stability of the cross-section and double spin asymmetry $A_{TT}$ at RHIC kinematics. We find that the QCD corrections
are non-negligible even at the level of asymmetries and that they strongly depend on the lepton kinematics.
Furthermore, we present two scenarios for transversely polarized parton distributions, based on the de Florian-Sassot-Stratmann-Vogelsang (DSSV) 
set of
longitudinally parton densities and fully evolved to NLO accuracy, that can be used for the evaluation of different observables involving transverse polarization.

\end{quote}

\end{titlepage}


\section{Introduction}

The partonic structure of polarized nucleons at the leading-twist (twist-2) level is characterized by the unpolarized, longitudinally polarized, and transversely~\cite{Ralston:1979ys,Artru:1989zv,Jaffe:1991kp,Jaffe:1991ra} polarized parton distribution functions $f$, $\Delta f$ and $\delta f$, respectively. Unpolarized parton distributions are known to a high degree of accuracy, allowing for very precise calculations at hadronic colliders, such as the LHC. On the other hand, regardless of much progress over the past three decades, many open questions 
concerning the helicity structure of the nucleon still remain. For 
example, we so far have only a rather unfinished picture of the individual longitudinal
polarizations of the light quarks and anti-quarks~\cite{dssvprd,dssvprl,deFlorian:2014yva}, and just a first hint on the 
helicity contribution of gluons inside the proton~\cite{deFlorian:2014yva,Nocera:2014gqa}. 
Nevertheless, a strong program of polarized pp collisions is now underway at the BNL Relativistic Heavy Ion Collider (RHIC)~\cite{rhicrev,spinplan}, aiming at further unraveling the spin structure of the proton.

Regarding the third leading-twist density, much less is experimentally known about the distributions of transversely polarized quarks in a transversely polarized proton 
(see \cite{Radici:2015mwa} for a recent extraction of valence transversity distributions from dihadron production).
 A number of different processes, including prompt photon, heavy flavour, inclusive hadrons and jet production have been proposed as observables to pin down the transversity distributions (see, e.g.,~\cite{Soffer:2002tf}).  It has long been recognized that {\it Drell-Yan} $Z$ boson production at the  RHIC may provide clean access to the transverse polarizations of quarks and anti-quarks in the colliding protons~\cite{Ji:1992ev,Vogelsang:1992jn,Martin:1997rz, Barone:1997mj,Martin:1999mg,  Soffer:2016oll}. The quantity of interest here is the double transversely polarized asymmetry defined as the ratio between the transversity cross section and the unpolarized one, as 
\beq
A_{TT}\equiv  \frac{d \sigma^{\uparrow \uparrow}
- d \sigma^{\uparrow\downarrow} - d \sigma^{\downarrow\uparrow} + d \sigma^{\downarrow\downarrow}}{d \sigma^{\uparrow \uparrow}
+d \sigma^{\uparrow\downarrow} + d \sigma^{\downarrow\uparrow} + d \sigma^{\downarrow\downarrow}}\equiv
\frac{d \delta \sigma}{d \sigma} \; ,
\label{eq:defasy}
\eeq
where the arrows indicate the corresponding transverse polarization of each beam.

During the last decades, a number of perturbative QCD next-to-leading (NLO) calculations became 
available for this observable,  either at the level of the fully inclusive 
cross-section~\cite{Vogelsang:1992jn,Contogouris:1994ws, Kamal:1995as} or
differential only on some of the variables~\cite{Vogelsang:1992jn,Martin:1999mg}.
For instance, the less inclusive analytical calculation presented in
Ref.\cite{Martin:1999mg} provides a result which is differential on the 
invariant mass and rapidity of the dilepton system and the azimuthal angle of
one lepton. But other relevant distributions, such as the transverse momentum of
each of the leptons, can not be reconstructed from those.

   While providing an estimate of the observables asymmetries,  this kind of 
   approach needs to be expanded in various 
ways. On one hand, there is an experimental issue: the detectors at RHIC
do not offer full coverage, which means that it is not always possible to 
reconstruct the momentum of the gauge boson from the leptonic final states. 
Furthermore, due to the acceptance of the detector and also in order to reduce the 
background, selection cuts are applied on several leptonic (and sometimes
hadronic) variables which are not described by more inclusive calculations.
On the other hand, at variance with the unpolarized and longitudinally 
polarized cases, for transverse polarization there is a strong azimuthal 
correlation between the spin of the protons and the momentum of the outgoing 
lepton. That makes indispensable to count with a fully differential description 
of the observable in terms of the leptons \footnote{Furthermore, in order to 
better reproduce the experimental cuts it is also necessary to be able to 
describe the hadronic activity in the final state}.
Indeed, the strategy adopted by the 
RHIC experiments is to detect the charged decay lepton and
determine its transverse momentum $p_{T}$ and rapidity $\eta$. 
The relevant process therefore becomes the reaction
$pp\to \ell^{\pm} X$, similar in spirit to the processes $pp\to \pi X$, $pp\to
{\mathrm{jet}}X$~\cite{pionref,deFlorian:1998qp,jetref} 
used at RHIC to determine gluon polarization in the
nucleon. 

The peculiar azimuthal dependence appearing in  transverse polarization scattering can be easily understood at the level of the matrix elements.
At the lowest order, the only partonic channel that contributes to the process is $ q(p_1) \bar{q}(p_2) \rightarrow e^-(l_1) e^+(l_2)$. The corresponding (color and spin averaged) transversely polarized matrix element is given by
\begin{equation}
\label{eq1}
\delta \overline{|{\cal M}|^2} = \frac{2}{3}  {\cal C} \left( 2  \frac{\hat{t}\hat{u}}{\hat{s}} s_1\cdot s_2  +\frac{4}{\hat{s}} s_1\cdot l_1 \, s_2\cdot l_1 \right)\, ,
\end{equation}
where $s_1\equiv (0; \cos \phi_1,\sin \phi_1,0)$ and $s_2\equiv (0; \cos \phi_2,\sin \phi_2,0)$ are the {\it transverse} spin vectors of the  incoming protons, and the usual Mandelstam variables are given by $\hat{s}=(p_1+p_2)^2$,  $\hat{t}=(p_1-l_1)^2$ and  $\hat{u}=(p_2-l_1)^2$. The {\it charge coefficient} ${\cal C}$,  adding contributions from both photon and $Z$-boson exchange, is given by
\begin{eqnarray}
\label{eq:c}
  {\cal C} \equiv e^4 e_q^2  + 2 e^2 e_q v_e v_q  \frac{\hat{s}(\hat{s}-M_Z^2) }{(\hat{s}-M_Z^2)^2+ \Gamma_Z^2 M_Z^2} - (v_e^2+a_e^2) (a_q^2-v_q^2 )  \frac{\hat{s}^2}{(\hat{s}-M_Z^2)^2+ \Gamma_Z^2 M_Z^2} \, ,
\end{eqnarray}
where, for the sake of simplicity in the notation, we write the corresponding weak coupling as proportional to $(v_i- a_i \gamma^5)$.

In the centre-of-mass frame of the incoming partons, the parenthesis in Eq.(\ref{eq1}) reads
\begin{equation}
 \frac{1}{2} \sin^2\theta \cos\left( 2\phi -\phi_1-\phi_2\right) \equiv \frac{1}{2} \sin^2\theta \cos( 2\Phi)\, ,
 \end{equation}
  where $\theta$ and $\phi$ are the polar and azimuthal angles of the lepton, respectively. As it occurs for other processes involving transversely polarized partons, the $ \cos( 2\Phi)$ term integrates to zero and, therefore, a special treatment is required to extract a non-vanishing asymmetry, as it will be discussed in Section \ref{secII}.

While leading-order (LO) calculations in hadronic collisions usually present only a {\it qualitative} description of an observable, higher order corrections are known to be large and needed to provide reliable {\it quantitative} predictions for a high-energy process. It is, therefore, crucial to determine the NLO QCD corrections. 

In general, the key issue here is to check the perturbative stability of the process considered, that is, to examine to which extent  the NLO corrections affect the cross sections and,  in spin physics the spin asymmetries relevant for experimental measurements. Only if the corrections are under control can a process that shows good sensitivity to a given transversity parton density be considered as a genuine probe for that, and be reliably used to extract accurate distributions from future data. 
Furthermore, the inclusion of extra partons in the NLO perturbative calculation also allows to improve the matching between the theoretical calculation and the realistic experimental conditions. This is particularly true when the calculation is performed at the fully differential level, such that all the four-momenta of all outgoing particles (leptons and partons) are available in order to apply the same cuts used at the experimental level. For that reason, we present here the first {\it fully differential} (in the hard cross section) NLO calculation for the production of single leptons, mediated by the exchange of a photon and a $Z$-boson, in collisions of transversely polarized protons $p \uparrow p\uparrow\to \ell^{\pm} X$.

The remainder of this paper is organized as follows:  in the next section
we very briefly discuss the non-standard characteristics of the NLO calculation with transverse polarization.  In Sec.~\ref{secIII} we introduce two different 
scenarios of transversely polarized distributions at NLO accuracy, a key ingredient for the calculation.
In Sec.~\ref{secIV} we study the perturbative stability of the different observables and provide the phenomenological
NLO results for the most relevant distributions and asymmetries. 
We finally conclude in Sec.~\ref{secV}.


\section{Next-to-leading Order Calculation \label{secII}}

\begin {figure}[!ht]
\begin{center}
\includegraphics[width = 5.in]{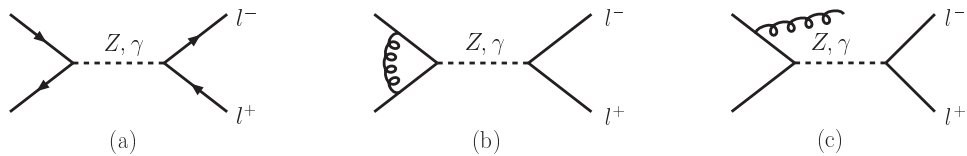} 
\end{center}
\caption{{\it Feynman diagrams for $Z,\gamma$ production with leptonic decay:
(a) leading-order, (b) NLO virtual correction, (c) NLO real emission. Crossed diagrams are not shown.}
\label{figWdiag} }
\end{figure}

In order to evaluate the NLO QCD corrections to the process we rely on the 
version of the subtraction method introduced and extensively discussed in 
Refs.~\cite{Frixione:1995ms,Frixione:1997np}, and later extended to the polarized case 
in Ref.~\cite{deFlorian:1998qp}. We refer the reader to those references for the details. 
The calculation is implemented in the Monte-Carlo like code `CHE' 
(standing for `Collisions at High Energies')
\footnote{The code is available upon request from deflo@unsam.edu.ar} which 
provides access to the full kinematics of the final-state particles,  allowing for
the computation of any infrared-safe observable in hadronic collisions and the 
implementation of realistic experimental cuts.
It is worth noticing that the same
code can compute the unpolarized, the longitudinally single polarized and the (longitudinally and transversely) double polarized
cross sections.
Even though the region of most interest at RHIC corresponds to the production of a lepton pair due to the decay of a $Z$ boson, 
the code also allows for the computation of the corrections arising from  photon exchange at the
same accuracy in perturbative QCD \footnote{Notice that the cross section is identically zero for transverse polarization in the case of $W^\pm$ exchange}. We show in Figure~\ref{figWdiag} some of the Feynman diagrams contributing at LO and NLO.

We point out that at NLO the contribution from photon exchange,
$q\bar{q}\to \gamma^* g$ followed by $\gamma^*\to \ell^+\ell^-$,
may generate large contributions when the high-transverse momentum
photon splits almost collinearly into the lepton pair, producing 
high-$p_{T}$ leptons with a very low invariant mass. A proper
treatment of this configuration would require the addition of a 
fragmentation contribution based on parton-to-dilepton fragmentation 
functions~\cite{Kang:2008wv}. On one hand, it is likely that
configurations with two nearly collinear leptons can be distinguished
experimentally from true single high-$p_{T}$ leptons. On the other hand, the kinematical region of interest
for transverse polarization at RHIC is limited to the high invariant mass configuration around the $Z$ mass, $M_Z$.
Therefore, in our calculation we can formally avoid such dangerous configurations by requiring 
the lepton pair to have an invariant mass $M_{{l_1}{l_2}}>10$ GeV, without any compromise in the result since most phenomenological analysis will 
actually demand $M_{{l_1}{l_2}} \gtrsim70$ GeV.

\vspace{1.7cm}

The Monte-Carlo like implementation relies on the integration by using numerical adaptive routines, such as Vegas, in order to improve the necessary cancellation of different terms in the subtraction method. The transversity cross section introduces an extra complication towards that,  due to the particular azimuthal dependence $\cos 2\Phi$ which integrates to zero over the full phase space.  In order to avoid that, and to produce results according to the conventional strategy, we multiply the corresponding squared matrix elements by  $\sign(\cos 2\Phi)$,
 such that the azimuthal integration becomes
\begin{equation}
\int_{-\pi}^{\pi}\cos 2\Phi \,  d\Phi \rightarrow  \left(   \int_{-\pi}^{-3\pi/4}  - \int_{-3\pi/4}^{-\pi/4}  +   \int_{-\pi/4}^{\pi/4}  - \int_{\pi/4}^{3\pi/4} + \int_{3\pi/4}^{\pi} \right) \cos 2\Phi \,  d\Phi  \, ,
\end{equation}
maximizing the transversity cross section.

As a check of the implementation of the calculation, we have also computed the fully inclusive transversely polarized cross sections, integrated over all lepton angles.  For these cross section, analytical results are  available~\cite{Vogelsang:1997ak}, with which we agree.

\section{Transversity parton distributions at NLO \label{secIII}}

In  analogy to the longitudinally polarized density $\Delta f$, the transversity distribution $\delta f$ is defined as the difference of finding a parton of flavor $f$ at a scale $Q$ with momentum fraction $x$ and its spin aligned $(\uparrow \uparrow)$ and anti-aligned  $(\downarrow \uparrow)$ to that of the transversely polarized nucleon:
\begin{equation}
\delta f(x,Q) \equiv f \uparrow \uparrow(x,Q) - f \downarrow \uparrow(x,Q) \, .
\end{equation}
At variance with the longitudinally polarized and unpolarized cases, there is no transversity gluon density for spin $1/2$ hadrons~\cite{Cortes:1991ja,Jaffe:1989xy}. 
The lack of a gluon distribution, and its corresponding mixing with quarks, has striking effects on the (factorization) scale dependence of the transversity densities, which evolve as {\it non-singlet} quantities. Valence and sea quark distributions evolve very similarly, with small differences that start at NLO accuracy.

In order to analyze the perturbative stability of the NLO cross section, it is indispensable to count with transversity parton distributions evolved with the corresponding NLO kernels~\cite{Vogelsang:1997ak,Hayashigaki:1997dn,Kumano:1997qp}. Given that little information on transversity distributions is available so far, we will present two {\it extreme scenarios} for them. The first one is based on Soffer's inequality ~\cite{Soffer:1994ww}
 \begin{equation}
2 |\delta f(x,Q)| \leq   f(x,Q)  + \Delta f(x,Q) \, ,
\end{equation}
which has been shown to be preserved under evolution at LO and NLO~\cite{Vogelsang:1997ak, Martin:1997rz,Bourrely:1997bx}. For the  {\it transversity maximally saturated} scenario we assume that the inequality is saturated (choosing the positive sign) at a low scale $Q=1$ GeV.  For the unpolarized distributions we use the MSTW set~\cite{Martin:2009iq}, while for the helicity densities we rely on the latest DSSV14~\cite{dssvprd,dssvprl,deFlorian:2014yva} analysis. By saturating the inequality at $Q=1$ GeV, one usually generates transversity distributions that can be {\it unnaturally} large, in particular in the sea quark sector. Given the non-singlet nature of the transversity distributions, the sea quark densities at $Q\sim M_Z$ can only be large at small $x$  if the same distribution is already sizable at the low initial scale $Q=1$ GeV. In contrast, the unpolarized and longitudinally polarized sea quark distributions are driven at small $x$ by their mixing to the gluon density through the evolution and can grow considerably.

\begin {figure}[!http]
\begin{center}
\includegraphics[width = 4.0in]{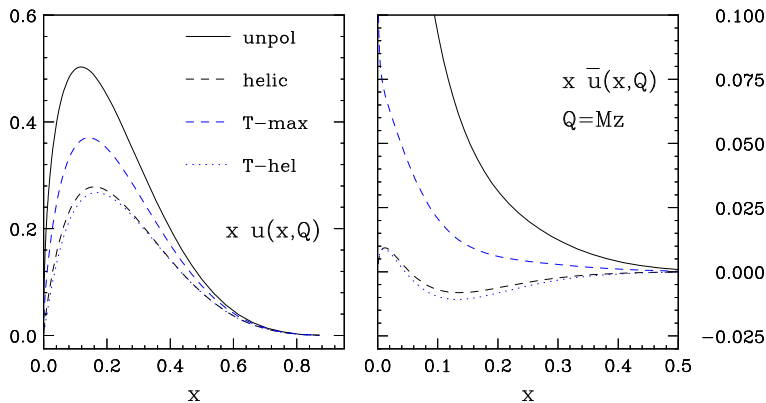} 
\end{center}
\caption{{\it Left: Next-to-leading order 
$x {u}(x,Q)$ evaluated at the scale $Q=91.2$~GeV for the unpolarized MSTW~\cite{Martin:2009iq} distributions (solid), helicity DSSV~\cite{dssvprd,dssvprl,deFlorian:2014yva} distribution (dashes), {\it transversity maximally saturated} (dashed-blue), 
and {\it transversity-helicity} distributions (dots-blue). Right: Same for $x \bar{u}(x,Q)$ (right-hand side).  } 
\label{figpdfu} }
\end{figure}

\begin {figure}[!http]
\begin{center}
\includegraphics[width = 4.0in]{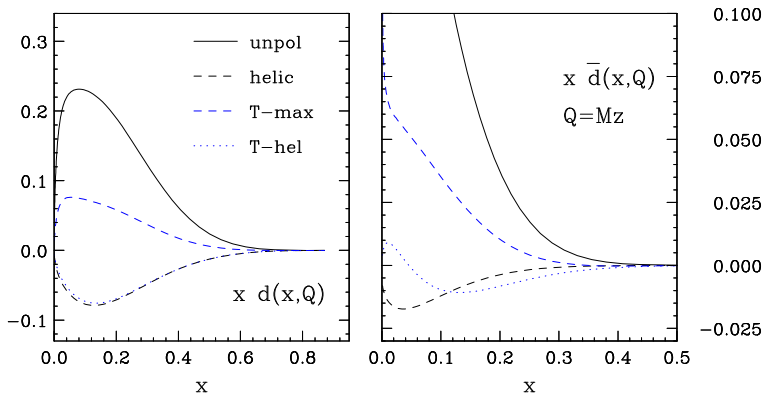} 
\end{center}
\caption{{\it Same as Figure~\ref{figpdfu} but for the  $x {d}(x,Q)$ and $x \bar{d}(x,Q)$ distributions.} 
\label{figpdfd} }
\end{figure}

A more conservative scenario relies on a possible analogy between longitudinally and transversely polarized quark densities. Since the assumption $\delta f(x,Q) \equiv  \Delta f(x,Q)$ can not be maintained for all scales $Q$ due to the different evolution of $\delta f(x,Q)$ and  $ \Delta f(x,Q)$, we set the equality between both distributions at the initial scale  $Q=1$ GeV  in the {\it transversity-helicity} scenario.
The result for both scenarios at $Q= M_Z$, along with the unpolarized and longitudinally polarized counterparts, are shown in Figs. \ref{figpdfu} and \ref{figpdfd} for the $u,\bar{u}$ and $d,\bar{d}$ quark distributions, respectively.

As can be observed, and in agreement with the arguments presented above, in the {\it transversity-helicity} scenario the quark densities follow the same trend of the helicity-distributions, while in the antiquark sector we see larger differences originated by the scale evolution. On the other hand, more sizable transversity distributions are obtained in the {\it transversity maximally saturated} scenario, where we also notice a notorious difference in the small $x$-behaviour between the transversity and unpolarized distributions due to their non-singlet and singlet nature, respectively.

Along this paper, we discard eventual contributions from heavy quark distributions in the polarized case, an rely only on the three massless flavor approach. Furthermore, we only produce NLO evolved parton densities, and use them as discussed in Section \ref{secIV} \footnote{A Fortran code with the sets of tranversely polarized parton distributions is available upon request from deflo@unsam.edu.ar}.


\section{Phenomenological Results for RHIC \label{secIV}}

In this section we analyze the perturbative stability of different observables in lepton production. 
We now use our NLO code to present some numerical results for polarized
$pp$ collisions at RHIC at center-of-mass energy $\sqrt{S}=500$ GeV. 
We do not include any QED or electroweak (EW) corrections, but choose the coupling constants $\alpha$ and 
$\sin^2\theta_W$ in the spirit of the `improved Born approximation'~\cite{IBA1,IBA2}, in order to effectively take into account 
the electroweak corrections. This approach results in $\sin^2\theta_W=0.23119$ and $\alpha=\alpha(M_Z)=1/128$. 
We also require the lepton pair to have an invariant mass $M_{{l_1}{l_2}}>10$ GeV, in order to avoid potentially 
large NLO contributions from production of a high-$p_T$ nearly real photon that subsequently decays into a pair of almost collinear leptons, as discussed
before.
We set the mass of the vector boson to $M_Z=91.1876$~GeV  and the corresponding decay width to 
$\Gamma_Z=2.4952$~GeV~\cite{PDG}. 
For the unpolarized cross section we will use the MSTW distributions with five massless flavors, while for the transversely polarized case we rely on the (3-flavor) sets
of distributions presented in Section  \ref{secIII}.

We study two different observables for lepton production in
$pp\to \ell^{-}X$: the 
transverse momentum ($p_{T}$) distribution of the electron 
with a rapidity cut of $|\eta_e|<1$, and the rapidity distribution with 
$p_{T}>20$ GeV. 
There are two hard scales in the process, which are of the same order:
the mass of the gauge boson and the transverse momentum of the observed 
lepton. We choose $\mu_F^2=\mu_R^2=(M_Z^2+p_{T}^2)/4$ as the 
default factorization and renormalization scales. We note that the scale
dependence of the cross sections and, in particular of the spin asymmetries
is extremely mild in case of vector boson production, so that other choices like
$\mu_F=\mu_R=M_Z$ or $\mu_F=\mu_R= M_{{l_1}{l_2}}$  provide rather similar results.

 Given that the main reason to study  polarized scattering is to shed light on the spin structure of the proton, and, in this particular case, to obtain information
 on the transversely polarized distributions, we begin by analyzing which is the sensitivity range of the observable in the momentum fraction carried by partons.
 With the selection cuts implemented in this analysis, the process is dominated by the kinematics on the $Z$-pole and, therefore, one expects a
 correlation between the partonic momentum fractions and the $Z$'s rapidity ($y_Z$), for which one has $x_{1,2}=\frac{M_Z}
{\sqrt{S}} e^{\pm y_Z}$ at the Born level.  It has been shown~\cite{deFlorian:2010aa,NadYuan2} 
that this relation between momentum fractions and rapidity at the gauge 
boson level is inherited by the lepton, even to NLO accuracy. 
A remarkably strong correlation is found between $\langle x_{1,2}\rangle$ 
and $\eta_e$ and, as a rough approximation, one can parameterize these 
correlations by the simple `empirical' formulas
\begin{equation}
\langle x_{1,2}\rangle \sim \frac{M_Z}{\sqrt{S}} e^{\pm \eta_e/2}\;.
 \end{equation}
Considering that RHIC experiments will allow to reach rapidities of the order of 
$|\eta_e|\sim 1$, one can expect sensitivity to the transversely polarized quark and anti-quark distributions in the region $0.07 \lesssim x \lesssim 0.4$.

\begin {figure}[!http]
\begin{center}
\includegraphics[width = 2.5in]{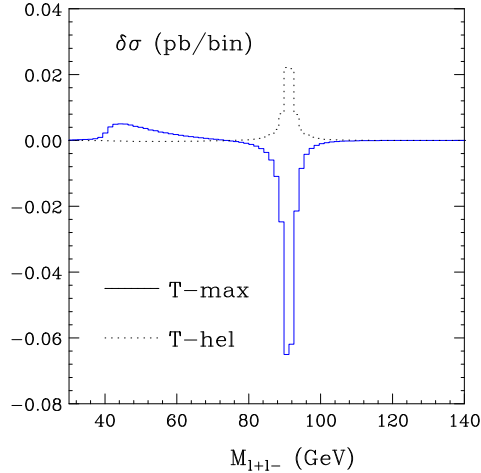} 
\end{center}
\caption{{\it Dilepton mass distribution for the transversely polarized cross section computed at NLO accuracy with the {\it transversity maximally saturated} (solid) and  {\it transversity-helicity} (dotted) polarized densities.} 
\label{figinv} }
\end{figure}

By observing the distributions in Fig.\ref{figpdfu}, it is clear that the leading $u,\bar{u}, d$ and $\bar{d}$ transversity distributions are always positive in that kinematical
range for the {\it transversity maximally saturated} scenario, while $\bar{u}$ and $\bar{d}$ densities are mostly negative (with a sign change in that relevant region) for the  {\it transversity-helicity} scenario. The overall sign of the transversely polarized cross section (and therefore the sign of the corresponding asymmetry) arises from the combination of the parton distributions and the partonic cross section. For $q\bar{q} \rightarrow \gamma^* \rightarrow e^- e^+$ annihilation the polarized partonic asymmetry is positive, after removing the overall $\cos 2\Phi$ term. The situation changes at the $Z$-pole due to the different ElectroWeak couplings, as observed at the leading order in Eq.(\ref{eq:c}), such that the ratio between the corresponding partonic contributions to the cross section is roughly given by
\begin{equation}
\label{eq:weak}
\frac{ \delta\sigma^{q\bar{q} \rightarrow  e^- e^+} (M_{{l_1}{l_2}}\sim M_Z)}{\delta \sigma^{q\bar{q}  \rightarrow e^- e^+} (M_{{l_1}{l_2}}\ll M_Z)} \sim - \frac{ (v_e^2+a_e^2)(a_q^2-v_q^2)}{e^4 e_q^2}
\end{equation}
and, therefore, the transversity partonic asymmetry becomes negative at $M_{{l_1}{l_2}}\sim M_Z$. We can observe this feature in Fig.(\ref{figinv}), where we present the dilepton invariant mass distribution for the transversely polarized cross section. 
The sign of the cross section around the peak is therefore fixed by (the opposite sign of) the one arising from the combination of the polarized parton distributions, 
resulting in a negative asymmetry for the  {\it transversity maximally saturated} scenario and a positive one for the {\it transversity-helicity} scenario (due to the mostly negative antiquark distributions). On the other hand, for invariant masses far from the $Z$ peak (i.e. $M_{{l_1}{l_2}}\lesssim 70$ GeV  or  $M_{{l_1}{l_2}}\gtrsim 110$ GeV ), the cross section is dominated by photon exchange and the opposite sign is observed.

We move now on the relevant issue of analyzing the perturbative stability of the QCD expansion for different observables. 
One usual way to display the size of radiative QCD corrections is in terms of a `$K$-factor', which represents the
ratio of the NLO and LO results. In the calculation of the numerator of
$K$ one obviously has to use NLO-evolved parton densities. As far as the
denominator is concerned, a natural definition requires the use of
LO-evolved parton densities. However, by using NLO-evolved parton
densities and LO partonic cross sections, one still obtains
a hadronic cross section accurate to LO, and therefore the
denominator of the $K$-factor can also be computed with
NLO-evolved parton densities. The longitudinally polarized parton distributions, which are at the basis of both transversity distribution scenarios, are not yet
as well determined as the unpolarized ones. Therefore, different results might arise for some of them when fits are performed at LO or at NLO
accuracy, resulting in rather large $K$-factors for the {\it distributions} themselves.  As an outcome of that, the use of LO distributions in the evaluation of the 
denominator could generate artificially large or small 
$K$-factors in the transversely polarized cross-sections, with effects far beyond those originated by the perturbative corrections. Therefore, along this paper,
we always use NLO distributions for both LO and NLO observables to emphasize the true outcome of the higher order terms.

\begin {figure}[!http]
\begin{center}
\includegraphics[width = 2.5in]{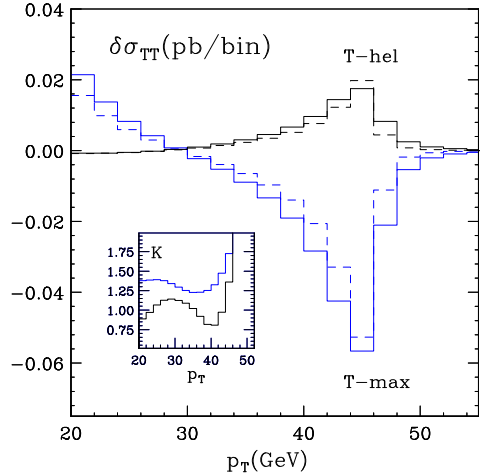} 
\end{center}
\caption{{\it Transverse momentum dependence of the NLO (solid) and LO (dashes) transversely polarized  cross sections. The corresponding $K$-factors are shown in the inset plot.} 
\label{figpt} }
\end{figure}

We start by presenting in Fig.(\ref{figpt})  the dependence of the transversely polarized  cross sections on the transverse momentum of the electron. On first hand, we observe that the cross sections are dominated by the production of leptons around the Jacobian peak $p_T\sim M_Z/2$. In this region the QCD corrections, as observed in the $K$-factors presented in the inset plot, become rather large and unstable. This is not unexpected: at LO, reaching $p_T> M_Z/2$ is only possible due to the finite width structure of the $Z$ boson, while starting at NLO  that region can be filled by the decay of leptons from a $Z$ boson with net transverse momentum, feature possible due to the emission of extra gluons at higher orders.
Therefore, that kinematical regime becomes very sensitive on soft gluon emission, and its proper description requires all-order resummation of the large logarithms that spoil the convergence of the perturbative expansion. 
However, after one integrates over a sufficiently large region of lepton transverse momentum, these logarithms turn into finite corrections and their resummation is not necessary. From the point of view of extracting transversely polarized parton distribution functions, it therefore seems advisable to focus on observables integrated over the lepton's transverse momentum (such as the rapidity dependence presented here), because these are insensitive to soft-gluon effects, and to use a plain NLO calculation.
On the other hand, at low transverse momentum, we observe a change of sign in the cross sections due to the dominance of the pure QED (photon-exchange) contribution, similarly to what occurs at low dilepton invariant mass, as already observed in Fig.(\ref{figinv}). In this kinematical region, due to the change of sign, the QCD corrections also become rather large and very much dependent on the scenario used for the transversity parton distributions.

\begin {figure}[!http]
\begin{center}
\includegraphics[width = 2.5in]{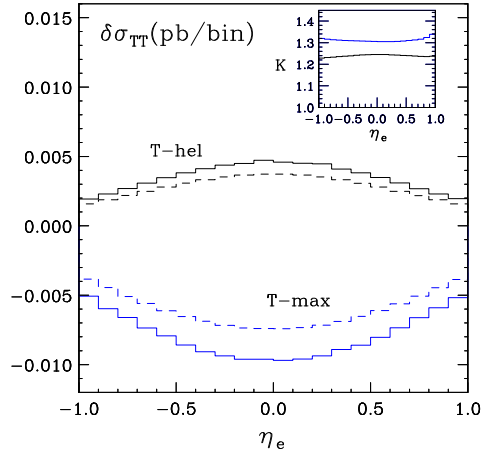} 
\end{center}
\caption{{\it Rapidity dependence of the NLO (solid) and LO (dashes) transversely polarized  cross sections. The corresponding $K$-factors are shown in the inset plot.} 
\label{figeta} }
\end{figure}

In Fig.(\ref{figeta}) we show the rapidity dependence of the NLO and LO transversely polarized  cross sections, for both scenarios of transversity parton distributions.  It is clear from there that the QCD corrections to the cross section are sizable and very much dependent on the set of distributions used. The inset plot displays the $K$-factors, explicitly manifesting corrections in the range of 20-35$\%$. It is important to notice that even in the rather restricted rapidity range relevant for RHIC, usual assumptions like {\it constant} $K$-factors, as those obtained from fully inclusive calculations of $Z$ production, would fail to provide an accurate description of this observable.

\begin {figure}[!h]
\begin{center}
\includegraphics[width = 2.5in]{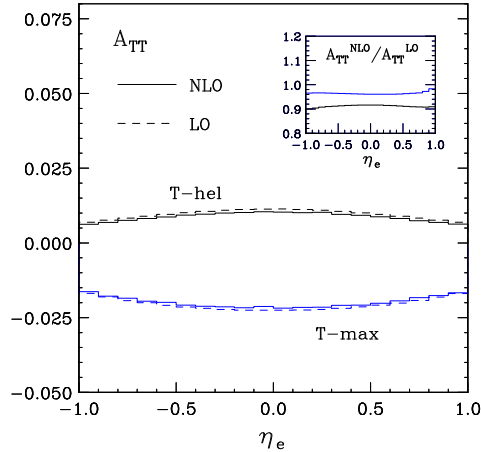} 
\end{center}
\caption{{\it  Rapidity dependence of the NLO (solid) and LO (dashes) transversely polarized  asymmetries for the two sets of the transversely polarized parton distribution functions defined above. The corresponding $K$-factors are shown in the inset plot.} 
\label{figasym} }
\end{figure}

Finally, we present in Fig.\ref{figasym} the corresponding LO and NLO results for the transversity asymmetry. For the sake of simplicity we do not include the {\it background} contribution that might arise 
from $pp \rightarrow W^{\pm} \rightarrow l^{\pm}\nu$ in the unpolarized cross section needed to define the asymmetry. That would only
result in a slightly smaller asymmetries, without any modification of the features presented along this paper, and, furthermore, is usually avoided by requiring the presence of two charged leptons in the detector \cite{Elke}.
The general features of the asymmetries can be easily understood.
 In the {\it transversity maximally saturated} we find negative (due to the $Z$ pole dominance) and larger  asymmetries than for the (positive asymmetry) {\it transversity-helicity} distributions, due to the more sizeable transverse polarization of both valence and quark densities in the first scenario.

It is also visible that, within the proposed scenarios, the asymmetries are at the few percent level, similarly to other observables involving transversely polarized beams~\cite{Soffer:2002tf}. In principle it would be possible to generate transversely polarized distributions with a larger polarization, assuming that the boundary condition is imposed at even lower initial scales, but that might turn out into rather unphysical scenarios. While large luminosities will be clearly needed at RHIC to perform the measurement, even the observation of the sign of the asymmetry would be of great help to improve our understanding on the spin content of the proton. For more details on the experimental possibilities for the measurement, we refer the reader to the recent analysis in~\cite{Soffer:2016oll}.

\begin {figure}[!h]
\begin{center}
\includegraphics[width = 2.5in]{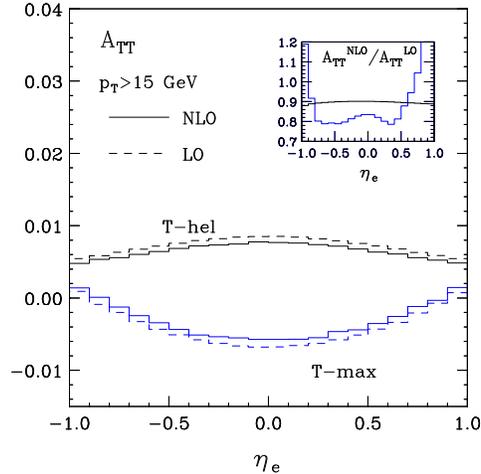} 
\end{center}
\caption{{\it  Rapidity dependence of the NLO (solid) and LO (dashes) transversely polarized  asymmetries for the two sets of the transversely polarized parton distribution functions, with a modified cut on the  transverse momentum of the electron $p_{T}>20$ GeV. The corresponding $K$-factors are shown in the inset plot.} 
\label{figasym15} }
\end{figure}

While asymmetries are in general rather stable under the QCD corrections, since many effects present in the individual cross sections cancel in the corresponding ratio, the NLO contributions still have a non-trivial impact. 
In the inset plot of Fig.\ref{figasym} we show the corresponding asymmetry $K_A\equiv \frac{A_{NLO}}{A_{LO}}$-factors, where we can observe corrections of the order of $10\%$ for the asymmetry computed with the {\it transversity-helicity} set. It is interesting to notice that $K_A$ is always below one for both sets of transversity distributions, but that this is not an overall feature of QCD. For example, a tiny modification in the transverse momentum cut for the lepton can produce a rather large effect in the observed asymmetries.
By lowering the corresponding cut from $20$  GeV to $15$ GeV,  as can be observed on the results presented in Fig.\ref{figasym15}, the asymmetries are considerably reduced with respect to the previous case and the NLO corrections become more sizable, with $K_A$ deviating  even further away from unity. This effect can be understood on simple basis; while the unpolarized cross section grows monotonically as the cuts become less restrictive, the transversely polarized cross section is reduced by a cancellation between the EW and photon contributions.
While a modification in the cut around 15-20 GeV does not affect substantially the pure EW term, which typically produces leptons with transverse momentum around $p_{T}\sim M_Z/2$, it does modify significantly the photon share that contributes  to the integral with the opposite sign and reduces the asymmetry. 
Furthermore, the size of the QCD corrections slightly depend on the relevant transverse momentum of the event: they are typically larger for {\it lower scale} contributions, such as those relevant for the QED part, than for {\it higher scale} ones, as those involve in the EW term.
This results in an even  more prominent cancellation between QED and EW contributions at NLO. Therefore, the perturbative stability of the asymmetry turns out to be affected also by the explicit leptonic cuts used in the analysis.

On the other hand, by selecting leptons with larger transverse momentum, or directly by choosing events with dilepton invariant mass  in the range $70$ GeV $ \le M_{{l_1}{l_2}} \le 110$ GeV one finds larger asymmetries with $K_A$ closer to one or even larger.
Therefore, it is clear that for a precise analysis of future RHIC data on this observable, a NLO {\it fully differential} calculation, such as presented here, is essential for a clear understanding of different observables even at the asymmetry level.

\section{Conclusions  \label{secV}}

In this paper, we have presented the first complete differential calculation
at next-to-leading order in perturbative QCD of the  Drell-Yan
cross section in transversely polarized hadronic collisions. The calculation is implemented in the Monte-Carlo like code 'CHE' 
 that already includes the unpolarized and longitudinally polarized cross sections.
Using the aforementioned code, we investigated in some detail
the phenomenological implications of jet production at RHIC
(polarized $pp$ collisions with a maximum centre-of-mass energy
of 500~GeV). We find that the QCD corrections are sizable, very much dependent on the
cuts and kinematic domain for the observed lepton,  and have a visible effect even for the transversely polarized double asymmetry.

\section*{Acknowledgments}
We are grateful to Werner Vogelsang for many useful discussions and Carla G\"obel for a careful reading of the manuscript.
This work has been partially supported by Conicet, ANPCyT and the von Humboldt Foundation. 
 We thank the Pontif\'\i cia Universidade Cat\'olica do Rio de Janeiro (PUC-Rio)  for the hospitality during the completion of this work.


\end{document}